# THE TELESCOPES AND PROCESSES
# OF THE AUSTRALIAN ASTRONOMICAL OBSERVATORY

ANDREW M. HOPKINS, GAYANDHI M. DE SILVA,
CHRIS M. SPRINGOB, STUART D. RYDER,
FRED G. WATSON AND MATTHEW M. COLLESS
*Australian Astronomical Observatory*
*P.O. Box 296*
*Epping NSW 1710, Australia*
ahopkins@aao.gov.au
gdesilva@aao.gov.au
springob@aao.gov.au
sdr@aao.gov.au
fgw@aao.gov.au
colless@aao.gov.au

**Abstract.** The Australian Astronomical Observatory operates the Anglo-Australian Telescope and the United Kingdom Schmidt Telescope in Australia, as well as coordinating access for the Australian community to the Gemini, Magellan, and other international telescope facilities. We review here the processes involved within the AAO related to allocating observing time on these facilities, as well as the impact on telescope use of both the Large Program projects and the AAO's instrument program.

## 1. Introduction

The Australian Astronomical Observatory (AAO)[1] is a division of the Australian Federal Government's Department of Innovation, Industry, Science and Research (DIISR). The AAO operates the Anglo-Australian Telescope (AAT) and the United Kingdom Schmidt Telescope (UKST) at Siding Spring Observatory, near Coonabarabran in north-western New South Wales. The AAO also hosts the Australian Gemini Office (AusGO)[2], which manages the allocation of time to Australian astronomers on the Gemini

---

[1]http://www.aao.gov.au/
[2]http://ausgo.aao.gov.au/

telescopes, as well as on Keck, Subaru and Magellan, through time-sharing or purchase agreements.

Here we outline the processes involved in allocating time on these facilities, as well as some reflections on the impact and role of Large Program projects at the AAT and the relationship between the AAO's instrumentation program and the science programs of our users.

## 2. The Anglo-Australian Telescope

The Anglo-Australian Telescope (AAT) is a 4 m optical telescope located at the Siding Spring Observatory, near the country town of Coonabarabran, NSW, Australia. It was commissioned in 1974 by the then Anglo-Australian Telescope Board. The AAT provides world-class observing facilities with a range of state-of-the-art instruments, which are constantly being upgraded to meet the demands of the scientific community. Observing time on the AAT is highly sought-after, by international observers as well as the Australian astronomical community.

The AAT instrument suite includes the 2dF/AAOmega multi-object fibre spectrograph, the SPIRAL integral-field spectrograph, the IRIS2 infrared imaging spectrograph, and the CYCLOPS/UCLES/UHRF high-resolution echelle spectrograph. The AAO is currently building the HERMES[3] multi-object high-resolution spectrograph, to be commissioned in late 2012, along with a range of technologically innovative new instruments[4] including the GNOSIS OH-suppression fibre-feed to IRIS2, and the SAMI multi-object integral-field spectrograph system using hexabundle fibre IFUs.

### 2.1. AAT TIME ALLOCATION OVERVIEW

Observing time is allocated on the AAT through the following process.[5]

**The Proposal Call.** AAT observing time is scheduled by semester, with A Semesters running from February to July and B Semesters running from August to January of the following year. Each year, a *Call for Proposals* is announced on the first of March and September for the coming B and A Semesters respectively, for which the proposal deadlines are 15th March and 15th September.

**Proposal Submission.** All observing proposals must be submitted via the AAT online proposal form. The form requests all critical data, those necessary for technical assessment, such as instrument details and observing restrictions, as well as those necessary for statistical time allocation, such as PI and Co-I affiliations, level of observing support, and students involved. While we have fixed page-limits for the Scientific Justification section of proposal, the format is at the discretion of proposers (we do, however, have guidelines stating that densely packed, small-font and unformatted text are unlikely to improve the chances of getting observing time). Three pages are allowed for normal programs, five pages for those requesting 'Long-Term' status, and ten pages for 'Large Programs'. Long-Term programs are those which require time distributed over several semesters, although the total time request need not be large. Large Programs are those requiring 50 nights or more (there is no set upper limit), usually, though not necessarily, extending over several semesters. To be competitive as a Large Program, the scientific goals must be groundbreaking and not just incremental. In addition, only five of the ten pages allowed for the justification are to be allocated to the scientific case. The remaining pages are to be used for the observing strategy (up to two pages), the project management plan (up to two pages), and a project timeline (up to one page). Other than adhering to the page restrictions, the proposers are free to format their Scientific Justification as they wish, and are then required to upload a PDF version into the online application form.

---

[3] http://www.aao.gov.au/HERMES/
[4] http://www.aao.gov.au/instsci/
[5] Details can also be found at http://www.aao.gov.au/astro/applying.html

**Technical Assessment and External Referees.** All AAT proposals submitted by the deadline are assessed for technical feasibility by AAO astronomers. The technical assessors remain anonymous, and do not contribute comments regarding the scientific merit of the proposal. The Technical Secretary collates these reports and is responsible for contacting the PI if there is a serious technical flaw. The Technical Secretary has the final say on whether a proposal is feasible, and on identifying an appropriate number of AAT nights to be allocated if awarded time.

Proposals classified as Large Programs, requesting over 50 nights of telescope time over one or more semesters, are sent to external referees to aid in assessing their scientific merit. The Technical Secretary is responsible for soliciting up to 3 referees per proposal, where the referees are experts in the proposed field of research. PIs will have the right of reply to matters raised by the referees, but the referees' identity remains strictly anonymous.

**Proposal ranking.** The Australian Time Allocation Committee (ATAC) is responsible for grading all AAT Observing proposals on their scientific merit. The ATAC members meet about 6 weeks after the proposal deadline. All proposal details and any supplementary material (e.g. referee reports) are provided to ATAC in advance of the meeting. At the meeting ATAC members discuss each proposal and assign a final grade (see next section on the grading procedure). Once the grading is complete, the Technical

Secretary prepares a draft schedule of the telescope allocations, which is then discussed with ATAC. In some cases (e.g. where two equally ranked proposals are at the cut-off) ATAC revises the relative ranking of proposals. Allocations made at the meeting are provisional only, and strictly confidential.

**Telescope Schedule.** The draft schedule and rankings are sent to the AAT Scheduler, who undertakes to publicly release the AAT schedule within one week of the ATAC meeting. This includes taking into account instrument setup nights, Director's Discretionary time and the roster of the support astronomers. Immediately before, or in conjunction with, the release of the AAT schedule, the ATAC Secretary emails all applicants giving the number of allocated nights, and general feedback, including the committee's reasons for the non-allocation of time.

## 2.2. AUSTRALIAN TIME ALLOCATION COMMITTEE

There are seven members of the Australian Time Allocation Committee (ATAC), chosen primarily from among the Australian community, and including at least one International member. In addition, a substitute member will also be appointed to take the place of any ATAC member who is unavailable for a meeting of the committee. Appointment to ATAC is usually for a 3-year term, although appointments may be for staggered terms to ensure a steady turnover in membership. Nominations to the committee will be periodically sought by the AAO Director. There will normally be no more than one representative on ATAC from any given institution.

**ATAC Meeting and grading procedure.** The committee meets twice each year, usually in the first or second week of May and November, to assess and rank proposals for the following semester. ATAC will normally meet in person in Australia, at the AAO in Sydney, but members unable to travel to Sydney are expected to participate via videoconference. Four members constitute a quorum. At the meeting the Chair (or in their absence, the Deputy Chair) presides. Allocations are made by a grading system (described below) carried out by the members present. ATAC members may also be required to vote on procedural or allocation matters. In the event of an equality of votes the Chair (or Deputy Chair, if presiding) has a casting as well as a deliberative vote.

In order to save time at the meeting, each panel member does a full science pre-grading of all the proposals beforehand, abstaining for those proposals in which a member is taking part. These 'pre-grades' are submitted to the ATAC Secretary before the meeting. Panel members are still free to change their votes at the meeting as subsequent discussion may change their opinions.

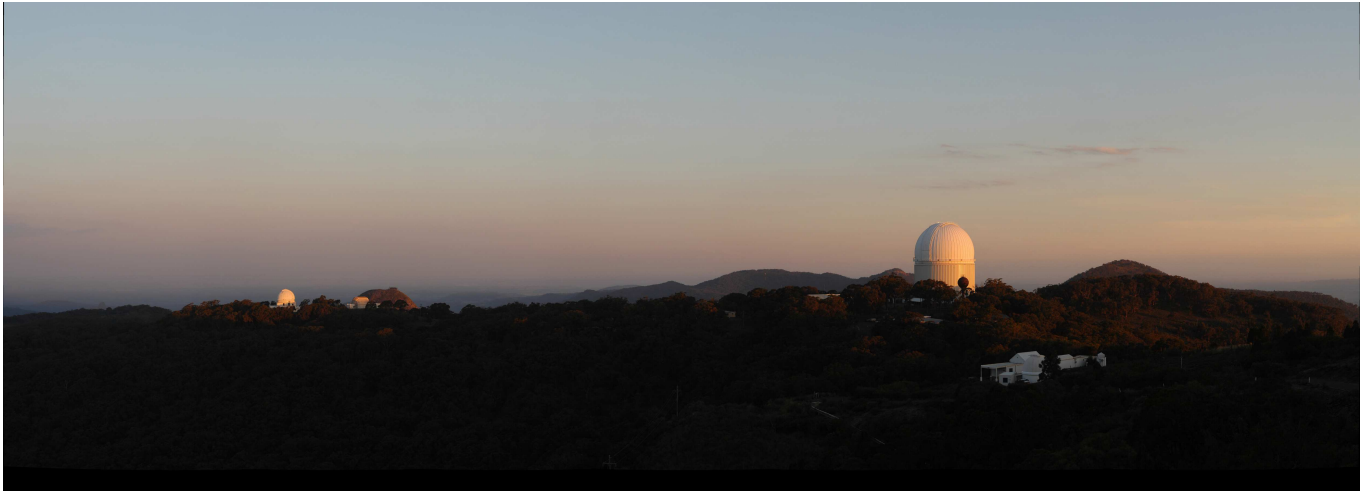

*Figure 1.* The AAT (right) and UKST (left) at Siding Spring Observatory. (Courtesy Fred Kamphues)

Committee members grade each of the proposals based on scientific merit, using the following guidelines: 5 = outstanding proposal; 4 = well above average proposal; 3 = good proposal; 2 = below average proposal; 1 = technically/scientifically defective proposal.

At the meeting, the proposals are discussed in order of their pre-grades, where the lowest pre-graded proposals are not further discussed, unless there is a sufficient dispersion in the pre-grades to warrant more investigation. The committee member assigned to a proposal gives a brief summary of the proposal, concluding with their scientific opinion of the application. The application is then discussed by the whole committee, and the final score is given in terms of the above grades. Fractional grades are permitted. Each ATAC member submits their final score anonymously to the Technical Secretary, who then determines the final averaged score.

Grades by committee members should be given on the scientific merit of the proposal, irrespective of whether dark, grey, or bright time is requested. Proposals are graded scientifically for the maximum number of nights/hours requested unless panel members feel the goals can be met in less time.

When a committee member is included in the list of applicants on a proposal, or otherwise feels they may have a conflict of interest, they must recuse themselves from the meeting during discussion and voting on that proposal.

**Responsibilities of committee members.** Each member of ATAC is expected to: (a) assess and grade each application prior to the meeting; (b) prepare a summary of the proposals allocated for presentation at the meeting; (c) collate a brief feedback statement for each proposal, based on discussion during the meeting; and (d) provide a point of contact for their constituents to communicate general issues with ATAC (although any potential matter of dispute arising from the meeting should be directed to the Chair).

In addition, the ATAC Chair (or in their absence, the Deputy Chair) is expected to: (a) oversee policy matters; (b) conduct the business at each meeting; (c) coordinate the allocation process at the meeting; (d) liaise with the ATAC Secretary and Technical Secretary over any matters arising or in development of new policies; (e) supervise the dispatch of feedback to all applicants after the meeting; and (f) liaise with the AAT Scheduler and Technical Secretary in the event of a scheduling conflict.

**The AAT Technical Secretary.** The AAT Technical Secretary is responsible for: (a) providing an updated list of available AAO instruments to the user community in advance of the application deadline; (b) supervising the receipt of ATAC applications via the online proposal form; (c) coordination of technical assessments by qualified AAO staff; (d) monitoring consistency of ATAC grades over consecutive semesters; (e) liaison with PIs prior to the meeting where there may be a serious technical problem with a proposal; and (f) providing a report to the AAO Director on over-subscription history, telescope usage, and updates on instrumentation status.

The Technical Secretary may not comment on scientific issues, unless invited to do so by the Chair, but is expected to attend the policy and scientific sessions of the ATAC meeting or at the very least be available for technical comment during the meeting.

## 2.3. ACCOUNTING FOR AAT OBSERVING TIME

The allocation of AAT time uses two parameters to balance rightful AAT share with high-quality science:

1. A specified fraction, $f_O$, of open-access time, taken out of Australia's (otherwise unconstrained) share, $f_A$, such that $f_A + f_O = 1$. The purpose of the open-access share is to foster international collaboration, as well as allowing high-ranking but internationally-dominated proposals to also use the AAT.

2. A super-majority threshold, M, used to measure the proportion of Australian involvement in a proposal. The purpose of the super-majority threshold is to ensure that time is awarded largely according to Australia's funding of the AAT's operation while still allowing (and even encouraging) some level of collaboration.

The time allocation procedure starts with ATAC ranking all proposals by scientific merit, without regard to the nationality of the applicants. The Technical Secretary then proceeds by:

— Initially drawing upon the Australian and Other time shares in proportion to the fraction of such proposers on each program. Note that nationality is determined by the location of the proposer's home institution, not by the proposer's citizenship. A proposal that is awarded $N$ nights and has $A$ Australians and $O$ Other proposers counts as $N_A$ Australian nights and $N_O$ Other nights, where $N_A = N \times A/(A + O)$ and $N_O = N \times O/(A + O)$. Then:

— If the Australian share is exhausted first, the remaining proposals are awarded time as ranked, regardless of nationality, or:

— If the Other share of AAT time is exhausted first, the remaining proposals are only awarded time from the residual Australian share if they (i) have an Australian PI and (ii) meet the super-majority criterion—i.e., if the fraction of Australians is greater than or equal to the super-majority threshold, $A/(A + O) \geq M$, rounding to the nearest whole percentage. The exception to this is when there are no qualifying pro-posals that can make use of the remaining time due to observational constraints.

Presently ATAC adopts a Australian fraction of 70%, and an Other fraction of 30%. The Australian super-majority threshold is 67%. These fractions reflect AAT demand (based on nationality) over past semesters and will be reviewed in the context of AAT demand over future semesters.

Once time has been allocated through a first pass of all eligible proposals, any remaining time will be filled through a second pass of the list. In this instance, proposals are taken solely on the basis of rank order, and the super-majority nationality criteria no longer apply. This means that time is not charged to partner shares nor repaid in future semesters. Rather, the nights are distributed from the remaining nights irrespective of which partner's share they come from.

Note that the actual fraction of time going to Other observers is generally higher than 30% (in fact typically around 40%) because there may be Other observers on proposals that are Australian-led and with an Australian super-majority.

The AAT Scheduler will, with guidance from the Chair of ATAC, make minor adjustments to the allocations to allow for practical matters such as dark/grey/bright time demand, scheduled instrument blocks, Director's time and so on.

## 2.4. AAT SERVICE OBSERVING

The AAO operates a service observing programme at the AAT for programmes that require up to six hours of observing time. Service time is normally allocated for programs that require a small amount of data to complete a programme, to look at individual targets of interest, or to try out new observing techniques.

The ATAC sets the total number of service nights each semester. This number is set so as to equalise the oversubscription rate between service proposals and regular proposals, and is typically 8–9 nights per semester, with at least one night per semester for each of the instruments.

A call for service proposals is issued three times a year, with deadlines on 1 February, 1 June, and 1 October. The proposals are reviewed by one internal and two external referees, on the basis of scientific merit. Proposals with an average referee grade of at least 2.5 (out of 5.0) are then added to the service queue, where they remain until either the observations have been conducted or 18 months have elapsed.

On each of the service nights, an AAO support astronomer conducts observations on one or more of the programs in the queue, selecting proposals to observe on the basis of the proposal's grade and the efficiency and fea-sibility of conducting the observations for the given proposal at that time of year. The data is then sent to the PI of the proposal in question, while the observations are logged, and the queue is updated.

## 2.5. LARGE PROGRAMS

The introduction of a Large Program model at the AAT has had an impact on the variety and scope of projects able to be facilitated through AAT observations. Examples of AAT Large Programs include the 2dF Galaxy Redshift Survey[6], the WiggleZ Dark Energy Survey[7], the Anglo-Australian Planet Search[8], and the Galaxy And Mass Assembly (GAMA) survey[9].

Each of these programs has involved substantial collaborations, and has resulted in high-profile and significant scientific results. The AAO expects Large Programs to be awarded in total at least 25% of the available time on the AAT, and in recent semesters as much as 50% of the time has been allocated to Large Projects. By facilitating these large allocations of time, and encouraging the community to work collaboratively on major projects, the impact of a 4 m class telescope is continuing to remain as strong as many 8–10 m class telescopes, and not far behind the most productive space-based observatories.

## 3. United Kingdom Schmidt Telescope

The UK Schmidt Telescope (UKST) is a survey telescope with an aperture of 1.2 m and a very wide field of view. The telescope was commissioned in 1973 and, until 1988, was operated by the Royal Observatory, Edinburgh. It became part of the AAO in June 1988.

Originally used exclusively for survey photography, the UKST was adapted for experimental multi-fibre spectroscopy in the 1980s, and a succession of prototype systems eventually culminated in the robotic 6dF multi-object spectrograph. This provides access to up to 150 targets over a six-degree diameter field of view. It was commissioned as a common-user instrument in 2001, replacing photography as the principal operational mode of the telescope.

Until 1 August 2005, UKST operations were funded by the AAO, with the telescope being used for 6dF Galaxy Survey[10] observations, as well as common-user non-survey observations. On that date, however, the current user-pays model was introduced, under which the AAO maintains the fa-cility and runs it as a cost-neutral asset. UKST users are required to pay the effective running costs for the period over which they make use of the telescope. In fact, with the exception of a very small number of Directors nights used for pilot observations, the sole user of the telescope since 2005 has been the international RAVE collaboration (the acronym stands for RAdial Velocity Experiment). The resulting RAVE[11] survey of stellar radial velocities and atmospheric parameters now comprises more than half a million spectra, which are being issued in a series of data releases.

---

[6] http://msowww.anu.edu.au/2dFGRS/
[7] http://wigglez.swin.edu.au/
[8] http://www.phys.unsw.edu.au/~cgt/planet/AAPS_Home.html
[9] http://www.gama-survey.org/
[10] http://www.aao.gov.au/local/www/6df/

The UKST is now owned by the Australian National University but is still operated by the Australian Astronomical Observatory. Once the RAVE survey is completed in mid-2012, collaborations interested in using the telescope will be invited to contact the AAO Director to discuss the details of proposed observing programs and the associated costs involved.

## 4. Allocation of time on external facilities

Australia has a 6.2% share in the Gemini Observatory partnership, providing about 15 nights per semester, split between the Gemini North and Gemini South 8 m telescopes. Access to a limited number of nights each semester on the Keck 10 m and Subaru 8 m telescopes is also offered by an exchange time agreement with Gemini. In addition, since 2007 Australia has purchased 15 nights per year through the Carnegie Institution for Science on the twin 6.5 m Magellan telescopes at Las Campanas Observatory to provide capabilities (e.g. high-resolution optical spectroscopy and wide-field optical imaging) not currently provided by the Gemini telescopes. The Australian Gemini Office (AusGO) within the AAO provides the interface between these facilities and Australian users, ranging from proposal submission to technical assessment, through to observation planning, assistance with data reduction, and publicising of results. The allocation of time each semester on these facilities is the responsibility of ATAC.

The process for allocating Australia's purchase of Magellan nights (nominally 8 nights in the A semester, and 7 nights in the B semester) is essentially the same as that described for the AAT, except that technical assessments are carried out by AusGO staff, and allocations are then forwarded to the Magellan Scheduler who makes best efforts to accommodate ATACs recommendations within pre-scheduled blocks of Carnegie Institution time on each Magellan telescope.

The process for allocating Australia's share of Gemini time is rather more complex. Technical assessments are carried out by AusGO staff prior to the ATAC meeting, and Principal Investigators given the chance to respond to any technical queries identified. After the ATAC meeting the ranked proposals and their recommended allocations are forwarded to the Gemini Observatory, which carries out an iterative queue merging process between the Gemini partner countries up to and during the International Time Allocation Committee (ITAC) meeting. The top ∼30% (by time allocated) of proposals go into queue Band 1, for which Gemini aim to achieve a 90% completion rate; the next 30% go into Band 2, which aims to have a 75% completion rate; the next 20% go into Band 3, for which 85% of programs which are started should receive at least 75% of their time; and the bottom 20% is available for 'Poor Weather' programs that can tolerate seeing $> 2''$ and/or $> 3$ mag of extinction by clouds. ATAC may award 'rollover' status for 2 more semesters to Band 1 programs (except for Target of Opportunity programs) so as to ensure their completion.

---

[11]http://www.rave-survey.aip.de/rave/

ATAC is required to forward programs which can use the full range of observing conditions and Right Ascensions, or risk forfeiting any unfilled time in that semester. During the ITAC meeting the ATAC Chair decides whether to support Joint proposals seeking some time from Australia, even if not all the other partners from which time is being sought have supported them with a comparable ranking. 'Classical' observing time on either of the Gemini telescopes, and on the Keck or Subaru telescopes via exchange time agreements, may be awarded by ATAC in units of 10 hours = 1 night, but these are top-sliced from ATACs total allocation, reducing the size of the queue bands accordingly. A complete description of the Gemini time allocation process can be found online[12].

## 5. Instrumentation

The AAO maintains a strong and innovative instrument science program for developing ground-breaking new instruments on both the AAT and other telescopes worldwide[13]. One of the impacts of having an instrumentation program that is integrated with the observer community has been the ability to be highly responsive to (i) community needs and requirements, and (ii) new technological developments, in particular in astrophotonics. One consequence has been the ability to develop highly innovative new instrumentation (such as the HERMES high-resolution multi-object spectrograph, as well as the GNOSIS OH-suppression fibre-feeds, and SAMI hexabundle IFU fibre-feeds) with a fast turnaround time. Importantly, this is also done in the environment of a strong relationship with the observer community, ensuring that high-impact scientific results flow rapidly from the commissioning and full-scale deployment of such new facilities.

---

[12]http://www.gemini.edu/sciops/observing-gemini/proposal-submission/tac-process
[13]http://www.aao.gov.au/astro/newinstrum.html

It is interesting to see how the science programs of the user community are both driving, and driven by, the instrumentation program at the AAO. In particular there is now a rapidly growing community in Australia with a focus on Galactic Archaeology that seeks to exploit the HERMES instrument.

## 6. Summary

This article has summarised the facilities offered by the Australian Astronomical Observatory and the procedures by which these resources are allocated under an open, competitive and peer-reviewed process. Both the facilities and the procedures have been developed over a long period in response to the scientific requirements of the user community (both Australian and international) and in an ongoing effort to optimise the scientific productivity and impact of the observatory. This evolutionary development has, both due to historical contingencies and by playing to emerging strengths, naturally led the AAO towards a specialisation in wide-field multi-object spectroscopy and large-scale survey programs. This particular focus exists, however, in the context a full-function telescope/instrumentation suite that caters to the very broad scientific needs of the entire community of optical/infrared astronomers in Australia. As those needs evolve in future, the AAO will develop new facilities and procedures so that it continues to serve its users and produce high-quality science.